\documentclass{emulateapj}
\slugcomment{Submitted to {\it ApJ} on 29 February 2008; accepted on 19 May 2008; published on 1 October 2008}
\shortauthors{Looper et al.}
\shorttitle{2MASS 1404AB: T Dwarf Binary}

\begin{document}

\title{Discovery of a T Dwarf Binary with the Largest Known 
$J$-Band Flux Reversal\altaffilmark{1}}

\author{Dagny L.\ Looper\altaffilmark{2}, 
Christopher R.\ Gelino\altaffilmark{3}, 
Adam J.\ Burgasser\altaffilmark{4},
J.\ Davy Kirkpatrick\altaffilmark{5}}

\altaffiltext{1}{Some of the data presented herein
were obtained at the W.M. Keck Observatory, which is operated as a 
scientific partnership among the California Institute of Technology, 
the University of California and the National Aeronautics and Space 
Administration. The Observatory was made possible by the generous
financial support of the W.M. Keck Foundation.  This paper includes data gathered with the 6.5-m Magellan Telescopes located at Las Campanas Observatory, Chile.}

\altaffiltext{2}{Institute for Astronomy, University of
Hawai'i, 2680 Woodlawn Dr, Honolulu, HI 96822}

\altaffiltext{3}{Spitzer Science Center, MS 220-6, 
California Institute of Technology, Pasadena, CA 91125}

\altaffiltext{4}{MIT Kavli Institute for Astrophysics \& 
Space Research, 77 Massachusetts Ave, Building 37-664B, Cambridge, 
MA 02139}

\altaffiltext{5}{Infrared Processing and Analysis Center, 
MS 100-22, California Institute of Technology, Pasadena, CA 91125}

\begin{abstract}

We present Keck laser guide star observations of two
T2.5 dwarfs -- 2MASS J11061197+2754225 and 2MASS J14044941$-$3159329 -- 
using NIRC2 on Keck-II 
and find 2MASS J14044941$-$3159329 to be a {0$\farcs$13} binary.  
This system has a secondary that is 0.45 mags brighter
than the primary in $J$-band but 0.49 mags fainter in $H$-band 
and 1.13 mags fainter in $K_s$-band.  
We use this relative photometry along with near-infrared synthetic 
modelling performed on the integrated light spectrum to 
derive component types of T1~$\pm$~1 for the primary and T5~$\pm$~1 
for the secondary.  Optical spectroscopy of this system obtained with
Magellan/LDSS-3 is also presented.  
This is the fourth L/T transition binary 
to show a flux reversal in the 1--1.2 $\mu$m regime and this 
one has the largest flux
reversal.  Unless the secondary is itself an
unresolved binary, the $J$-band magnitude difference between the
secondary and primary shows that the $J$-band ``bump'' is indeed a 
real feature and not an artifact caused by unresolved binarity.  

\end{abstract}


\section{Introduction}

As brown dwarfs cool and pass through the L/T spectral class boundary, 
their spectral morphologies transition from red
near-infrared (NIR) colors of the L dwarf class, caused by condensate dust in
their photospheres, to blue NIR colors of the T dwarf class,
where their photospheres are relatively clear of dust.  This transition is
rapid, as implied by the nearly flat effective temperature 
scale around 1400 K for NIR L7--T5 dwarfs 
(\citealt{2004AJ....128.1733G,2005ARA&A..43..195K}, 
bottom panel of Fig.\ 7 in that paper).  Within this transition, a remarkable
brightening in $J$-band ($\Delta$M$_J\sim$~1) 
from spectral types $\sim$T1$-$T5,
known as the $J$-band ``bump'' 
\citep{2002AJ....124.1170D,2003AJ....126..975T,2004AJ....127.2948V}, 
has been noted.  

Two withstanding mechanisms to explain this brightening have 
been suggested:  
(1) the ``patchy clouds'' model -- proposed by
\cite{2002ApJ...571L.151B} (see also \citealt{2001ApJ...556..872A}),
suggesting that the break-up of clouds in the atmosphere allows hot flux
from inner layers to emerge (analogous to the 5 $\mu$m hot spots of
Jupiter), and (2) the ``sudden downpour'' model 
-- proposed by \cite{2004AJ....127.3553K}, suggesting
that the dust clouds suddenly condense out due to an increase in
sedimentation efficiency.  

Recent studies of brown dwarf binaries have revealed that a fraction 
of the amplitude in this bump 
can be explained by systems appearing overluminous due to 
binarity (i.e., ``crypto-binarity'';
\citealt{2006ApJS..166..585B,2006ApJ...647.1393L,2006ApJ...640.1063B});
however, some part of this brightening is intrinsic to the atmospheres
as they cool.  This was revealed by HST/WFPC2 imaging of 
2MASS J17281150+3948593AB (hereafter 2MASS 1728AB), the first
1--1.2 $\mu$m flux reversal binary, which has a 
T dwarf secondary brighter in $z$-band but fainter in $i$-band
than the mid-to late-type L dwarf primary \citep{2003AJ....125.3302G}.  
No $J$-band resolved photometry for this system has been published.
Two later discoveries of 1--1.2 $\mu$m flux reversal binaries -- SDSS
J102109.69$-$030420.1 (hereafter SDSS 1021; HST/NICMOS, 
\citealt{2006ApJS..166..585B}) 
and SDSS J153417.05+161546.1 (hereafter SDSS 1534; Keck LGS AO/NIRC2, 
\citealt{2006ApJ...647.1393L}) -- 
provided additional information on the flux reversals.  Both systems 
were found to have a secondary brighter than the primary 
in $J$-band (see Table 1).  

L dwarf/T dwarf binaries such as those listed above provide crucial
information on the L/T transition, as the components of these systems
are likely coeval, with common ages and compositions.  In an effort to
identify additional systems, we have performed high angular resolution
imaging of two T2.5 dwarfs 
-- 2MASS J11061197+2754225 (hereafter 2MASS 1106) and 2MASS
J14044941$-$3159329 (hereafter 2MASS 1404).  These objects were identified
in a NIR proper motion survey (\citealt{2007AJ....134.1162L}; 
Kirkpatrick et al$.$, in prep) based on multi-epoch data
from the Two Micron All Sky Survey (2MASS; \citealt{2006AJ....131.1163S}).
Both systems are in the field, unassociated with any 
higher mass stars within a radius of 10 
arcminutes\footnote{See http://simbad.u-strasbg.fr/simbad/.}.
2MASS 1404 was discovered to be a
fourth resolved L/T transition system showing this 1--1.2 $\mu$m flux 
reversal.  These observations were made
using the Keck II sodium laser guide star adaptive optics 
system.  We describe these observations and optical spectroscopy 
of 2MASS 1404 in $\S$2,
discuss the results of this imaging and implications for the $J$-band
bump in $\S$3, and give our conclusions in $\S$4. 

\section{Observations}

\subsection{High Resolution Imaging: Keck II 10.0-m/NIRC2 LGS AO}

We used the Keck II 10.0 m Telescope\footnote{Located on the summit of 
Mauna Kea, Hawaii.} Sodium Laser Guide Star Adaptive Optics (LGS AO) system 
\citep{2006SPIE.6272E...7W,2006PASP..118..310V} on 2006 June 3 UT 
to observe 2MASS 1106 and 2MASS 1404.  
Observations were taken using the NIRC2 narrow camera and the Mauna Kea
Observatories (MKO) filter set \citep{2002PASP..114..180T}: 
$K_s$-band (2.15 $\mu$m) for 2MASS 1106 and $J$ (1.25 $\mu$m), $H$ 
(1.635 $\mu$m), and $K_s$-bands for 2MASS 1404.  
The field of view was 10$\farcs$2 with a 
0{\farcs}00994~pixel scale.  Nearby ($<$~60$\arcsec$) and bright
($R<19$ mag) stars were selected from the USNO-B1.0 catalog 
\citep{2003AJ....125..984M} to provide for 
tip-tilt (TT) sensing.  For 2MASS 1106, the TT star (USNO-B1.0 1179-0233699) 
had a brightness of $R=15.2$ mag and a separation of $\sim$52$\arcsec$.  
For 2MASS 1404, the TT star (USNO-B1.0 0580-0365843) 
had a brightness of $R=16.3$ mag and a separation of $\sim$29$\arcsec$.  
Conditions were clear and photometric with good seeing at the start of
the night (0$\farcs$7 at $K$-band) with slight degradation afterwards.

For 2MASS 1106, we dithered by a few arcseconds between positions 
to obtain a set of
eight images with an integration time of 60 sec for each position, for a
total on-source integration time of 480 sec.  For 2MASS 1404, we
dithered by a few arcseconds between positions, for a total of 
six images for each of the three filters.  Integration times per 
position were 
90 sec, 60 sec, and 60 sec for a total integration time of 540 sec, 360
sec, and 360 sec in the $J$, $H$, and $K_s$ filters,
respectively.  The full widths at half maximum (FWHM) in the $K_s$ filter
were 0$\farcs068$ and 0$\farcs065$ for 2MASS 1106 and 2MASS 1404,
respectively.  

We employed standard reduction techniques.  Flats were created
from differenced lights-on and lights-off images of the telescope dome
interior.  A super-sky frame was 
created from the median of the individual science frames and 
was subsequently subtracted from each frame.  The background subtracted 
images were shifted so that the
target landed on the same position and were then stacked to form the
final mosaic.

\begin{figure}
\epsscale{0.38}
\plotone{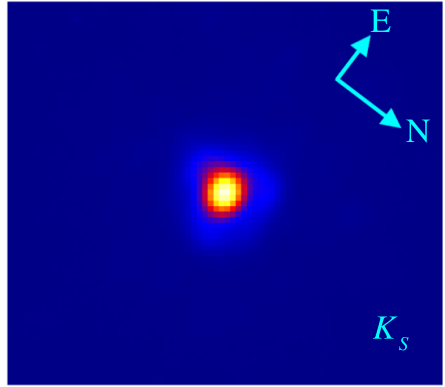}
\caption{Keck LGS AO image of 2MASS 1106 in $K_s$ band.  
North and East are indicated by an arrow.  
The image is $\sim$0$\farcs8$ on a side, with no other sources besides
the target detected in the full-sized ($\sim$13$\farcs5$) field-of-view.
The FWHM is 0$\farcs$068. 
\label{fig1}}
\end{figure}

\begin{figure}
\epsscale{1.0}
\plotone{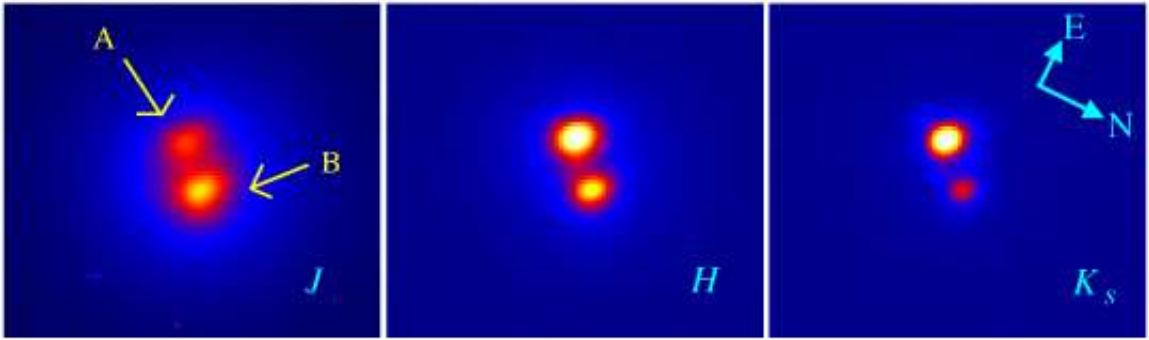}
\caption{Keck LGS AO images of 2MASS 1404AB in $J$, $H$,
and $K_s$ bands.  North and East are indicated by an arrow 
and is the same for all images.  
Note that the B component is brighter in $J$-band than
the A component but is fainter in both $H$- and $K_s$-bands.  Each
subimage is $\sim$0$\farcs$8 on a side, and the separation of the
binary is 0$\farcs$1336$\pm$0$\farcs$0006 with a position angle of
311.8$^\circ$~$\pm$~0.7$^\circ$.  No other sources besides components A and B were detected in the full-sized ($\sim$14$\arcsec$) field-of-view.
The FWHM is 0$\farcs$065 in $K_s$-band.
\label{fig2}}
\end{figure}

In the mosaic of 2MASS 1106 (see Fig.~\ref{fig1}), 
the target is not elongated and no
other objects, companions or background sources, are seen. 
On the other hand,     
we have resolved 2MASS 1404 into two components, shown in
Fig.~\ref{fig2}.  Based on the large proper motion of this system
($\mu$~=~0$\farcs$35 yr$^{-1}$) and the elapsed 5.3 yrs between the first
2MASS observation of this field (2001 Feb 4 UT) and our observation with
Keck II (2006 Jun 3 UT), we should be able to see both components 
separated in the 2MASS image, which has a plate scale of
$\sim$1$\arcsec$, if they were not physically associated.  Because we do
not see an object on the 2001 2MASS image that is positionally coincident
with the double seen in our 2006 Keck image, we conclude that this is a
physical binary unresolved in 2MASS images.  
Using the $K_s$-band final mosaic, which has the smallest FWHM, we find 
a separation, $\rho$, of 0$\farcs$1336~$\pm$~0$\farcs$0006 and a position
angle, $\phi$, of 311.8$^\circ$~$\pm$~0.7$^\circ$.

We performed an image subtraction technique in all three bands to obtain 
final flux ratios for the 2MASS 1404AB system. To do this, we computed the 
centroid for the component to be
subtracted and placed this component at the center of a 257$\times$257 
pixel subimage.
This subimage was copied, rotated 180 degrees, and then subtracted
from the original, non-rotated subimage. The quality of the subtraction 
was checked both visually
and by examination of the radial profile of the unsubtracted component. 
In many cases, small offsets
had to be applied by hand to obtain optimal subtraction. Aperture 
photometry was
performed on the unsubtracted objects and used to obtain the flux ratio 
difference of the components.

Errors in the rotation-subtraction technique were estimated by creating 
fake binaries from (presumably)
single objects observed with NIRC2 in LGS AO mode and having similar 
image qualities as 2MASS 1404AB. The
fake binaries were created at 25 different separations and 
position angles.  The same
rotation-subtraction technique was performed on these binaries as 
was done for 2MASS 1404AB.  The
magnitude of each component was taken to be the average of the 25 
separate measurements and the error was
the standard deviation of the measurements. 

The final errors in the photometry for 2MASS 1404AB are the square root 
of the sum of squares for the
following eight terms: the errors in the single point-spread function (PSF) photometry (both A 
and B), 
the standard deviation of the binary PSF photometry (both A and B), the 
difference between the
single and binary PSF photometry (both A and B), and the errors in 2MASS 
1404AB photometry (both
A and B).  The final flux ratios for 2MASS 1404AB are shown in Table 2.

\subsection{Red Optical Spectroscopy: Magellan 6.5-m/LDSS3}

Red optical spectroscopy ($\sim$6000--10500 $\AA$) of the 2MASS 1404AB
system was obtained on 2006 May 8
(UT) with LDSS-3 (upgraded from LDSS-2, \citealt{1994PASP..106..983A}) 
on the 6.5-m Magellan Clay Telescope.  Conditions were
clear with good seeing ($\sim$0$\farcs$6 at $R$-band).  
The VPH-red grism (660 lines/mm) and the OG590 longpass filter were 
used with a 0$\farcs$75 (4 pixels) slit, 
resulting in R~=~$\lambda/\Delta\lambda~\approx$~1800.  The slit
was rotated to the parallactic angle to minimize slit losses.
Dispersion across the chip was 1.2 \AA/pixel.  We obtained two
exposures of 1500 sec each, for a total integration time of 50 
minutes at an average air mass of 1.01.  The flux standard star LTT 7987 
\citep{1994PASP..106..566H} was observed the
previous night (2006 May 7 UT) using an identical set-up.  Calibration exposures were taken using the HeNeAr arc lamp and the flat-field quartz lamp.
The G2 V star HD 127526 was observed immediately prior to 2MASS 1404AB to use as a correction for telluric absorption.  

LDSS-3 data were reduced in the IRAF\footnote{IRAF is
distributed by the National Optical
Astronomy Observatories, which are operated by the Association of
Universities for Research in Astronomy, Inc., under cooperative
agreement with the National Science Foundation.} environment.
Raw science images
were first trimmed, bias-subtracted
and then divided by the
normalized, median-combined and bias-subtracted set of flat field frames.  Spectra were optimally extracted using the APALL task, with the extraction
of HD 127526 used as a dispersion template for 2MASS 1404AB.
Wavelength solutions were determined from the arc lamp spectra;
solutions were accurate to $\sim$0.1 pixels, or $\sim$0.12~{\AA}.
Flux calibration was determined
using the tasks STANDARD and SENSFUNC with
observations of LTT 7987, adequate over the
spectral range 6000--10000 {\AA}.
Corrections to telluric O$_2$ (6860--6960 {\AA} B-band,
7580--7700 {\AA} A-band)
and H$_2$O (7150--7300 and 9270--9675 {\AA}) absorption bands
were computed using the G dwarf spectrum.

\begin{figure}
\epsscale{1.0}
\plotone{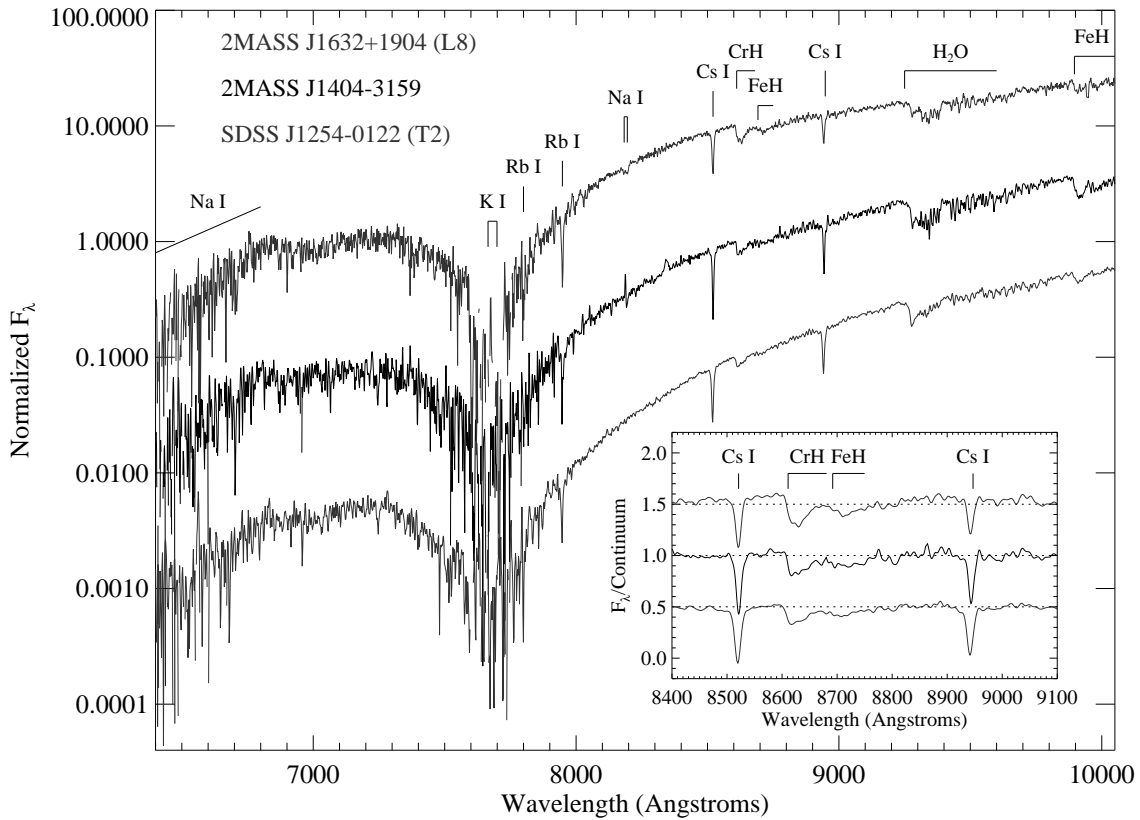}
\caption{Red optical spectrum of 2MASS 1404AB (obtained with
LDSS-3/Magellan; optical T0~$\pm$~1) 
in comparison to 2MASS 1632
(L8 optical standard; \citealt{1999ApJ...519..802K}) and SDSS 1254 
(T2 optical standard; \citealt{2003ApJ...594..510B}) normalized and shown 
on a log flux scale.  The insert shows the region between 8400--9100
\AA, where the spectra have been divided by a 4th-order polynomial fit
to the continua.  2MASS 1404AB and SDSS 1254 have been smoothed down to
approximately the same resolution as 2MASS 1632 (R~$\approx$~1000).  Major
atomic and molecular features are labeled, and the spectra are separated
along the vertical axis for clarity.  2MASS 1404AB appears to be
intermediate in type between the L8 and T2 optical standards.
\label{fig3}}
\end{figure}

The optical spectrum is shown in Fig.~\ref{fig3}, along with the L8 optical
standard 2MASS J1632291+190441 (hereafter 2MASS 1632, 
\citealt{1999ApJ...519..802K}) 
and the T2 optical standard SDSS J125453.90$-$012247.4 (hereafter SDSS 1254, 
\citealt{2003ApJ...594..510B}) for comparison.  The spectra are shown on
a log flux scale and have been smoothed to R$\approx$1000 to match the
resolution of 2MASS 1632 observed with Keck/LRIS.  Overall, the continuum
of 2MASS 1404AB is intermediate between the L8 and T2 standards, consistent
with an optical spectral type of T0.  The inset box in this same figure shows the Cs I lines and FeH/CrH bands between 8400
and 9100 \AA.  The spectra shown in this insert have been divided
through by a 4th-order polynomial fit to the continuum in order to
highlight the strength of these features.  The Cs I lines are of
comparable strength in 2MASS 1404AB as in SDSS 1254, while the FeH/CrH band
in 2MASS 1404AB is intermediate between the L8 and T2 standards.  We
therefore adopt an optical spectral type of T0~$\pm$~1 for the 2MASS 1404AB 
system.  

\section{Analysis}

\subsection{New M$_{JHK_s}$-Spectral Type Relations Based on 2MASS Photometry}

\begin{figure}
\epsscale{1.0}
\plotone{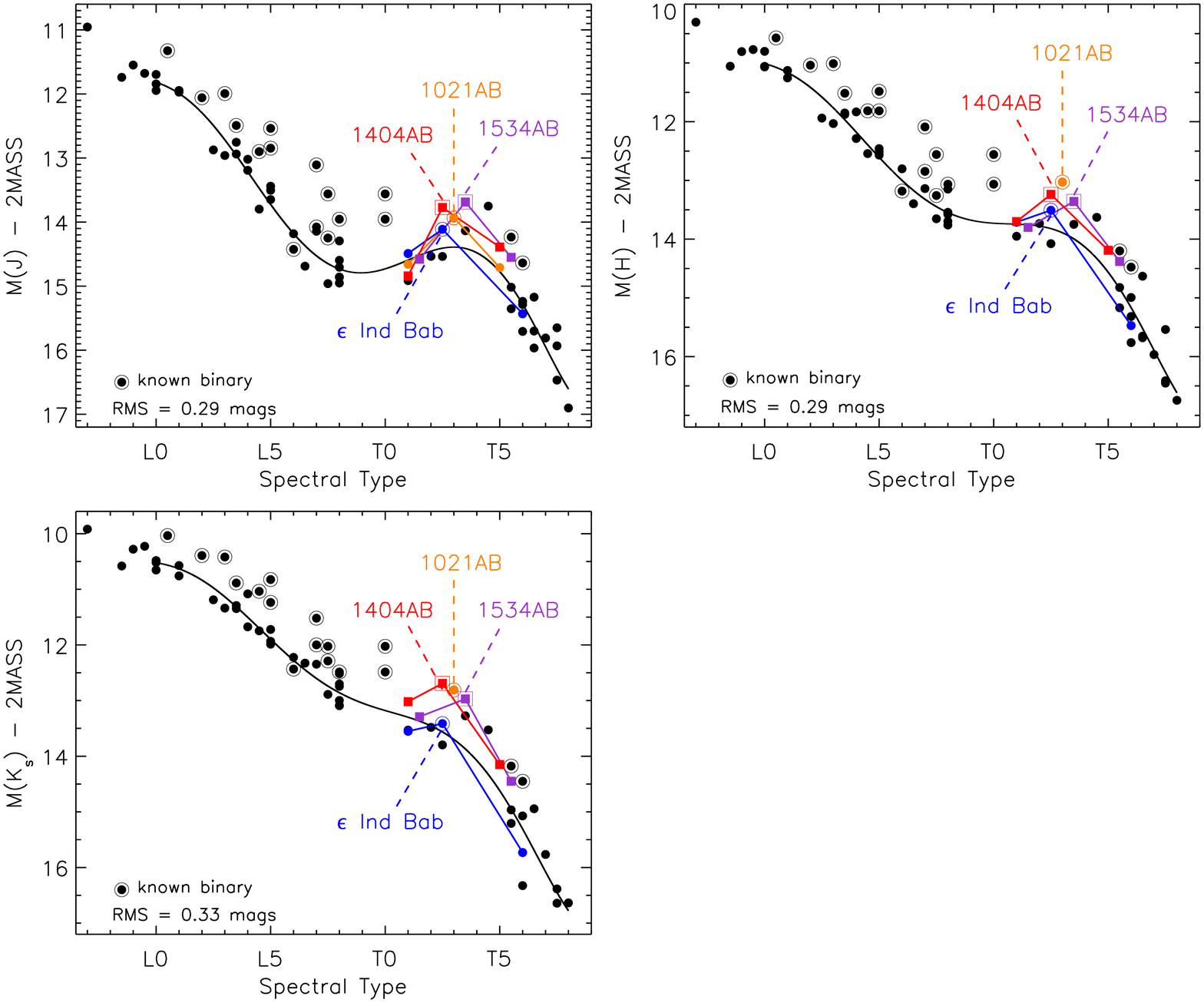}
\caption{M$_{JHK_s}$-vs-SpT of optical L dwarfs (on
\citealt{1999ApJ...519..802K} scheme) and NIR T dwarfs (on
\citealt{2006ApJ...637.1067B} scheme) with parallax measurements of SNR~$>$~5
(solid points).  The (unweighted) sixth-degree 
polynomial fits defined in $\S$3.1 are shown as solid lines.  
Late type M dwarfs were also included in these 
fits to prevent an artificial downturn at L0.  
Known binaries are encircled and were not included in the fit, with the
exception of $\epsilon$ Ind Bab (blue), where the spectroscopically 
classified components have been used.  
\textit{Top:} A clear brightening in M$_J$
is seen from $\sim$T1--T5, known as the $J$-band bump.  The three
known $J$-band flux reversal binaries -- SDSS 1021AB, SDSS 1534AB, and 2MASS
1404AB -- and their A+B components are shown as orange, purple, and red,
respectively.  2MASS 1404AB and SDSS 1534AB are indicated by squares, not
circles, since no parallax measurement has been made for either of these
sources.  No $H$- and $K_s$-band resolved photometry is available for 
SDSS 1021AB.  
\label{fig4}}
\end{figure}

We have derived $JHK_s$ absolute
magnitude versus spectral type relations using optically classified
late-type M and L dwarfs and
NIR classified T dwarfs with parallax measurements of signal-to-noise ratio (S/N)~$>$~5 and not known
to be binary (collected from: 
\citealt{1997A&A...323L..49P,2002AJ....124.1170D,
2003AJ....126..975T,2004AJ....127.2948V}).  
The coefficients of the sixth-degree polynomial fits to this 
unweighted data are given in Table 3 and shown graphically in 
Fig.~\ref{fig4}.  
We have used optical spectral classification for the L dwarfs because 
no formal L dwarf NIR spectral classification scheme has been
constructed\footnote{\cite{2002ApJ...564..466G} have identified 
classification indices based on
NIR molecular absorption bands.}.  These relations use the new
NIR spectral classification scheme for T dwarfs \citep{2006ApJ...637.1067B},
superceding previous M$_{JHK}$ versus spectral type (SpT) 
relations on the old NIR T dwarf scheme 
\citep{2002ApJ...564..466G,2002ApJ...571L.151B}.  

\subsection{Component Spectral Types of 2MASS 1404AB}

Since the relative $J$- and $H$-band photometry is on the MKO photometric
system and the relative $K$-band photometry is on the 2MASS photometric
system, we need to convert the $K$-band photometry onto the same MKO
system.  To do this, we examined a set of 25 L6.0--T2.0 dwarfs,
described below, and
computed the (2MASS$-$MKO) $K$-band color term as $-$0.01~$\pm$~0.04
averaged over these spectral classes.  We show the breakdown of this
color term per spectral class in Table 4.  Since this color term
difference is near zero, we assume the two are equivalent and proceed 
to determining the component MKO photometry, using the following steps:  

\begin{enumerate}

\item Convert $J_{AB}$ (2MASS) to $J_{AB}$ (MKO) using the 
    (MKO$-$2MASS) color term computed from the spectrum of 2MASS 1404AB
    \citep{2007AJ....134.1162L}. $J_{AB}$ (MKO$-$2MASS)~=~$-$0.17; $J_{AB}$
    (MKO)~=~15.41~$\pm$~0.07. 

\item Decompose the system:  
    $J_A$~=~$J_{AB}$ + 2.5~log(1 + 10$^{0.4~\times~\Delta J}$); 
    $\sigma_{J_A}^2$~=~$\sigma_{J_{AB}}^2$+$\frac{\sigma_{\Delta
    J}^2}{(10^{-0.4~\times~\Delta J}-1)^2}$, where 
    $\Delta J$~=~$J_A-J_B$.
    $J_A$ (MKO)~=~16.46~$\pm$~0.12.

\item Similarly, $J_B$~=~$J_{AB}$ + 2.5~log(1 + 10$^{-0.4~\times~\Delta J}$);
    $\sigma_{J_B}^2$~=~$\sigma_{J_{AB}}^2$+$\frac{\sigma_{\Delta
    J}^2}{(10^{0.4~\times~\Delta J}-1)^2}$, where 
    $\Delta J$~=~$J_A-J_B$.  $J_B$ (MKO)~=~15.93~$\pm$~0.09.
\end{enumerate}

In the same manner, 
this procedure was carried out to determine the MKO $H$-band and $K$-band 
component photometry (see Table 5).

To estimate NIR spectral types 
for the AB components, we compared the integrated light spectrum to 
a suite of synthetic IRTF/SpeX spectra constructed in the same manner as 
\cite{2008arXiv0803.0295B}.  The constituents for these synthetic spectra are
the 25 L6$-$T2 dwarfs, used for the primaries, and the 38 T3$-$T8
dwarfs, used for the secondaries, as listed in 
\cite{2008arXiv0803.0295B}\footnote{Also, see 
http://web.mit.edu/ajb/www/browndwarfs/spexprism/html/binarytemplates.html
for a list of these 63 spectra, where we have eliminated 2MASS
J21513839$-$4853542 (T4) and 2MASS J05103520$-$4208140 (T5) from the
list of T3$-$T8 dwarfs used due to the poor signal-to-noise ratio (SNR) 
of their spectra.}, which is
drawn from \cite{2004AJ....127.2856B,2006ApJ...637.1067B, 
2006ApJ...639.1095B,2007ApJ...659..655B,2006AJ....131.2722C,2004ApJ...604L..61C,
2007ApJ...655..522L,2007AJ....134.1162L,2006ApJ...639.1114R,
2007AJ....133.2320S}.  
In cases where more than one spectrum for an object was available, 
we used the highest SNR spectrum available.  All combinations of primary
and secondary were scaled together using the M$_J$ magnitude from the
M$_J$-SpT relation derived above in $\S$3.1.
This resulted in 1000 
synthetic spectra.  To estimate the goodness of fit between the data 
(2MASS 1404AB) and each model, we first interpolated the flux of the
models onto the wavelength scale of the data and normalized each to their peak
flux over the 1.2--1.3 $\mu$m window.  We then calculated the $\chi^2$ values\footnote{Defined as $\chi^2$~$\equiv$~$\Sigma_{\lambda}$
($\frac{f_{\lambda}(1404AB)-f_{\lambda}(M)}{\sigma_{\lambda}})^2$, where
$f_{\lambda}(1404AB)$ is the data for 2MASS 1404AB, $f_{\lambda}(M)$ is
the synthetic model spectrum, and $\sigma_{\lambda}$ are the errors in
the data for 2MASS 1404AB.  The summation is performed over the
wavelengths 0.95--1.35, 1.45--1.8, and 2.0--2.35 $\mu$m.}
between the data and each model over the wavelength ranges 0.95--1.35 $\mu$m, 
1.45--1.8 $\mu$m, and 2.0--2.35 $\mu$m to avoid low S/N
regions of the spectra.  In addition, we calculated $\chi^2$ for a range
of multiplicative scale factors (0.5 to 1.5) of the data, in steps of
0.01, to eliminate the normalization bias, selecting the normalization
that yielded $\chi^2_{min}$. 

\begin{figure}
\epsscale{1.0}
\plotone{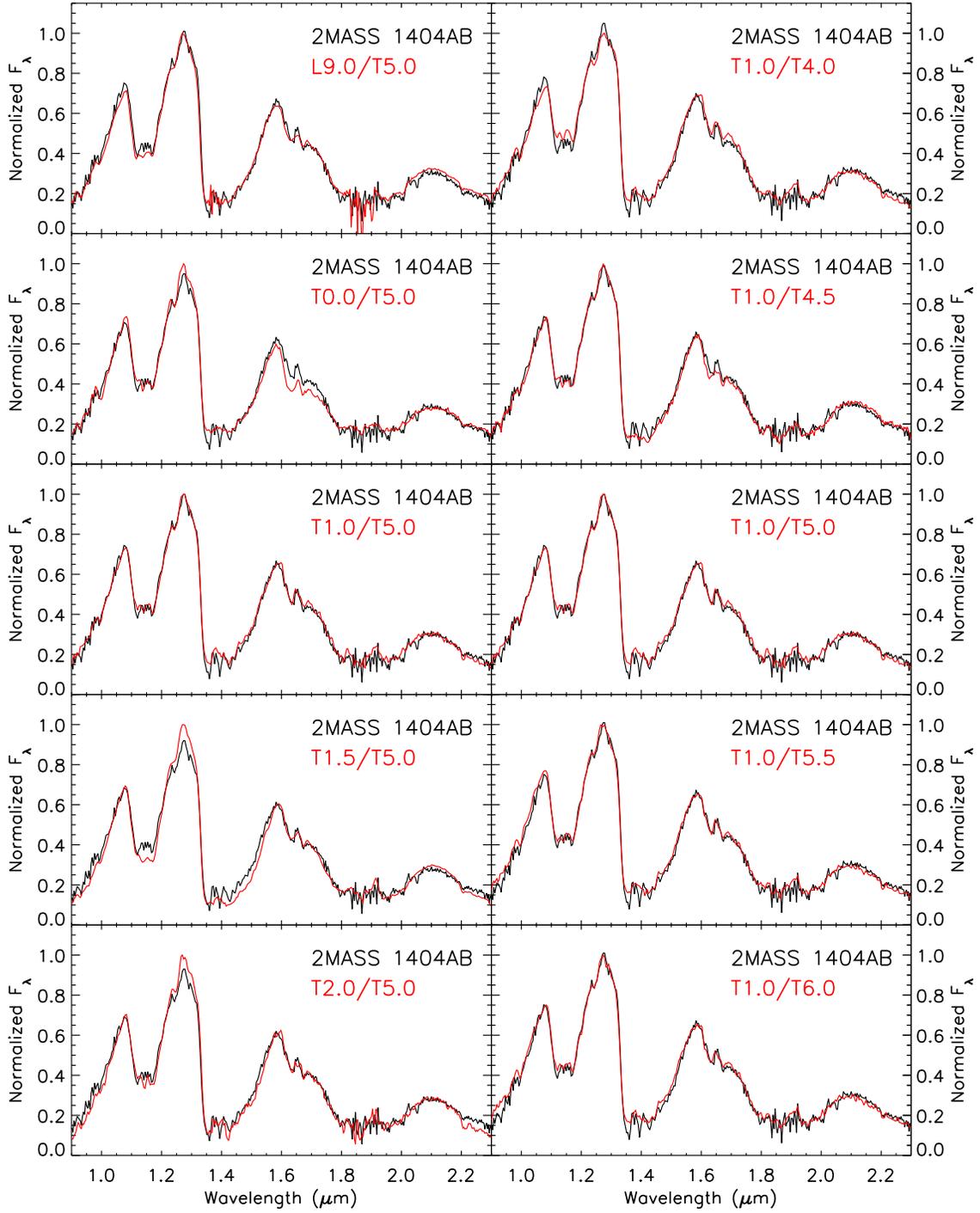}
\caption{Comparison of the T2.5 integrated light spectrum of 
2MASS 1404AB (black; \citealt{2007AJ....134.1162L}) with the best synthetic 
spectral fits (red; see $\S$3.2).  The T1 and T5 combination provides
the best fit across all wavelengths to the spectrum of 2MASS 1404AB.
\label{fig5}}
\end{figure}

From the 1000 model spectra, we selected the 15 spectra with the smallest
$\chi^2$ values for visual inspection.  The model spectrum created
from SDSS J205235.31$-$160929.8 (T1, hereafter SDSS 2052; 
\citealt{2006AJ....131.2722C}) 
for the primary and from 2MASS J23312378$-$4718274 (T5, hereafter 2MASS 2331; 
\citealt{2004AJ....127.2856B}) for the secondary with the same normalization 
provided the best fit and yielded 2MASS $\Delta J$~=~0.18 mag.  
We also examined model combinations 
constructed from two earlier type (L9 and T0) and two later type (T1.5 and T2)
primaries with a T5 used as the secondary.  In addition, we examined
model combinations constructed from a T1 as the primary and two
earlier type (T4 and T4.5) and two later type 
(T5.5 and T6) secondaries.  Since many
spectra with identical spectral types were available, we used 
those with model spectra that minimized $\chi^2$.  
All fits are shown in Fig.~\ref{fig5}.  The best fit is clearly
the T1/T5 model, which shows an excellent match to the data of 2MASS
1404AB over all wavelengths.  Deviating the spectral type 
of either the primary or secondary by a subtype or more 
while holding the other fixed, results in a degraded goodness
of fit to the data.  We therefore estimate a final NIR spectral type of
the primary as T1~$\pm$~1 and of the secondary as T5~$\pm$~1. 

Using the MKO component photometry of 2MASS 1404AB and component
spectral type estimates of T1 and T5, we compute the 2MASS component
photometry (see Table 5) 
using the appropriate transformations from \cite{2004PASP..116....9S}.
These relations are dependent on spectral type with errors that add
negligibly to the component photometric errors.
We can compare the relative 2MASS $J$-band photometry obtained 
from imaging, $\Delta J$~=~0.45~$\pm$~0.15, to that obtained 
from our synthetic spectral modeling, $\Delta J$~=~0.18,
using a T1 primary and a T5 secondary.  The relative 2MASS
$J$-band photometry obtained from our synthetic spectral modeling is 1.8 sigma from our relative 2MASS $J$-band photometry obtained from
imaging.  It confirms that the secondary is indeed brighter than the
primary in $J$-band.

\begin{figure}
\epsscale{0.7}
\plotone{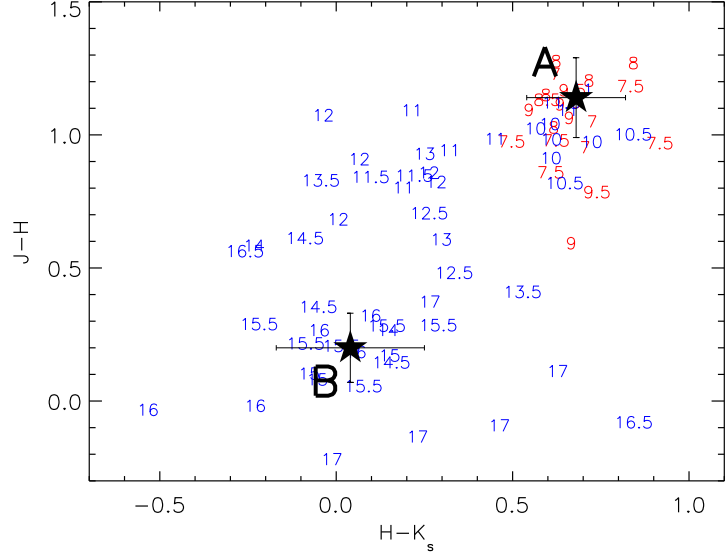}
\caption{NIR color-color plot of 74 known NIR L7-T7 dwarfs with
$\sigma_{J-H}$ and $\sigma_{H-K_s}<$~0.3 mags and spectral types with
uncertainties $\le$~1 subtype.  L dwarfs
are shown in red with spectral types represented by numbers $-$ 7~=~L7, 
8~=~L8,
etc.  T dwarfs are shown in blue with spectral types represented by numbers
$-$ 10~=~T0, 15~=~T5, etc.  The resolved colors of 2MASS 1404AB are
shown as stars with associated error bars.
\label{fig6}}
\end{figure}

As a check on our spectral type estimates, we compute 2MASS NIR colors of each
component and compare these to the 2MASS NIR colors of published L and T 
dwarfs\footnote{See http://dwarfarchives.org.} in Fig.~\ref{fig6}.  
In this color-color diagram, component A is seen coincident with a cluster of
L7$-$T1 dwarfs.  Component B is associated with a 
cluster of T4$-$T6 dwarfs.  This color comparison yields 
an average spectral type estimate of $\sim$L9 for component A and 
$\sim$T5 for component B.  

\begin{figure}
\epsscale{0.9}
\plotone{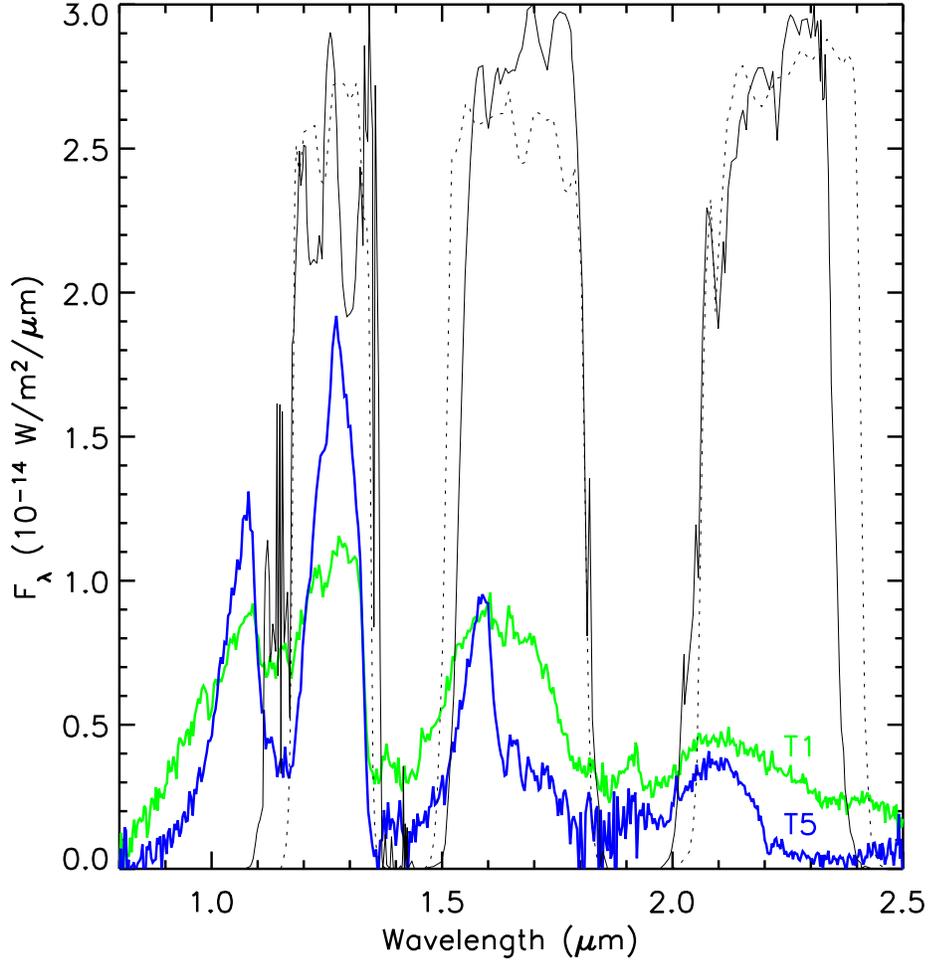}
\caption{\textit{Left}: SDSS J205235.31$-$160929.8 (NIR T1; 
\citealt{2006AJ....131.2722C}), shown in green, and  
2MASS J23312378$-$4718274 (NIR T5; \citealt{2004AJ....127.2856B}), shown in
blue, scaled 
by the MKO magnitudes of the A and B components of 2MASS 1404AB.
The 2MASS $JHK_s$
(solid lines) and MKO $JHK$ (dotted lines) transmission plus atmospheric
profiles are overlaid.  
\label{fig7}}
\end{figure}

In Fig.~\ref{fig7}, we plot SDSS 2052 and 2MASS 2331 scaled with the 
relative photometry of the A and B components of 2MASS 1404AB, with the 2MASS
and MKO filter transmission profiles overlaid.  
The 2MASS $J$-band filter profile extends to
slightly shorter wavelengths, covering more of the CH$_4$ absorption
band at $\sim$1.15 $\mu$m, than does the MKO
$J$-band profile.  This explains the reduced relative 2MASS 
$J$-band photometry of
the two components compared to the relative MKO $J$-band photometry, 
since the secondary (T5) has larger CH$_4$ absorption than does the 
primary (T1).  The flux peaks, however, are more illuminating than the
broadband photometry.  
The peak at 1.25 $\mu$m of the scaled T5 is
$\sim$70\% brighter than that of the scaled T1, and the 1.05 $\mu$m peak 
(outside the $J$-band filter profile) is $\sim$40\% brighter.  
This redistribution of flux into the 1.05 and 1.25 $\mu$m 
peaks is remarkably similar to
that seen for SDSS 1021AB and SDSS 1534AB.  
SDSS 1021B is $\sim$30\% and $\sim$25\%
brighter at the 1.25 and 1.05 $\mu$m bands than SDSS 1021A,
respectively.  Likewise, SDSS 1534B is $\sim$30\% and $\sim$20\%
brighter at the 1.25 and 1.05 $\mu$m bands than SDSS 1534A.
While no atmospheric models can accurately reproduce this flux
redistribution, we refer the reader to \cite{2006ApJS..166..585B} 
for a qualitative description of how this brightening may arise.

\subsection{Construction of Color-Magnitude Diagrams}

Using the new M$_{JHK_s}$-SpT relations derived in $\S$3.1, 
we estimate distances 
for component A, assuming a T1 spectral type, of $\sim$25.8, $\sim$22.6, 
and $\sim$20.2 pc, respectively, yielding a mean distance of $\sim$23 pc. 
We can now place 2MASS 1404AB and its components on a
color-magnitude diagram (see Fig.~\ref{fig4}).  Likewise, SDSS 
1021AB and SDSS 1534AB are also shown, and these 
have been converted onto the 2MASS photometric system using the spectral
type estimates of their components, their component MKO photometry, and
the appropriate transformations from \cite{2004PASP..116....9S}.  
The parallax measurement for SDSS 1021AB \citep{2003AJ....126..975T}
is used to place it on the diagram.  Since SDSS 1534AB lacks a
parallax measurement, we use the M$_{JHK_s}$-SpT relations for component
SDSS 1534A (d$_{est}\approx$~41 pc) to place SDSS 1534AB on the diagram.  

Of note, only two other T dwarfs in the $J$-band bump, 
SDSS J175032.96+175903.9 (T3.5; hereafter
SDSS 1750) and 2MASS J05591914$-$1404488 (T4.5; hereafter 2MASS 0559), 
lie above the $J$-band fit, by $\sim$0.3 and $\sim$0.9 mag, respectively.  
Neither object has been resolved in high-angular resolution studies
\citep{2006ApJS..166..585B,2003ApJ...586..512B}, nor show any signs of
unusual metallicity or gravity effects in their spectra.  
These objects could still be
binaries with very small separations and/or imaged at an unpropitious
place in their orbits, as was originally the case with Kelu-1
\citep{2005ApJ...634..616L,2006PASP..118..611G}.  However, 
\cite{2007ApJ...666.1205Z} have monitored 2MASS 0559 over a period of 
4.37 years with radial velocity measurements and have ruled out
companions more massive than 10 M$_{Jup}$ in orbits of $\sim$1 yr and 
companions more massive than 2 M$_{Jup}$ in orbits of less than a few days.  

In order to eliminate the $J$-band bump, SDSS 1750 would have to be an 
unresolved binary, 2MASS 0559 would have to be an unresolved triple 
(since $\Delta J$~$>$~0.75 mags) with the above restrictions on the
orbital periods of its components, and 2MASS 1404AB would also have to 
be a triple system instead of a binary.  The small observed frequency of 
brown dwarf triple systems (3$^{+4}_{-1}$\% -- not considering
selection effects; \citealt{2007prpl.conf..427B}) suggests this is highly 
unlikely.  Hence, the amplitude of the $J$-band bump is probably 
at least $\sim$0.5 mag as illustrated by the components of 2MASS 1404AB 
and likely as high as $\sim$1 mag in light of the recent null result of 
radial velocity companions to 2MASS 0559 by \cite{2007ApJ...666.1205Z}.  
A parallax measurement and resolved spectroscopy of this system are
needed to accurately place the components on a color magnitude diagram.  
Radial velocity monitoring of the 2MASS
1404AB system can resolve if this brightening is 
intrinsic to the atmospheres as they cool or if higher-order
multiplicity is responsible for this peculiar flux reversal at 1--1.2
$\mu$m. 

\subsection{Physical Parameters of 2MASS 1404AB}

\begin{figure}
\epsscale{0.9}
\plotone{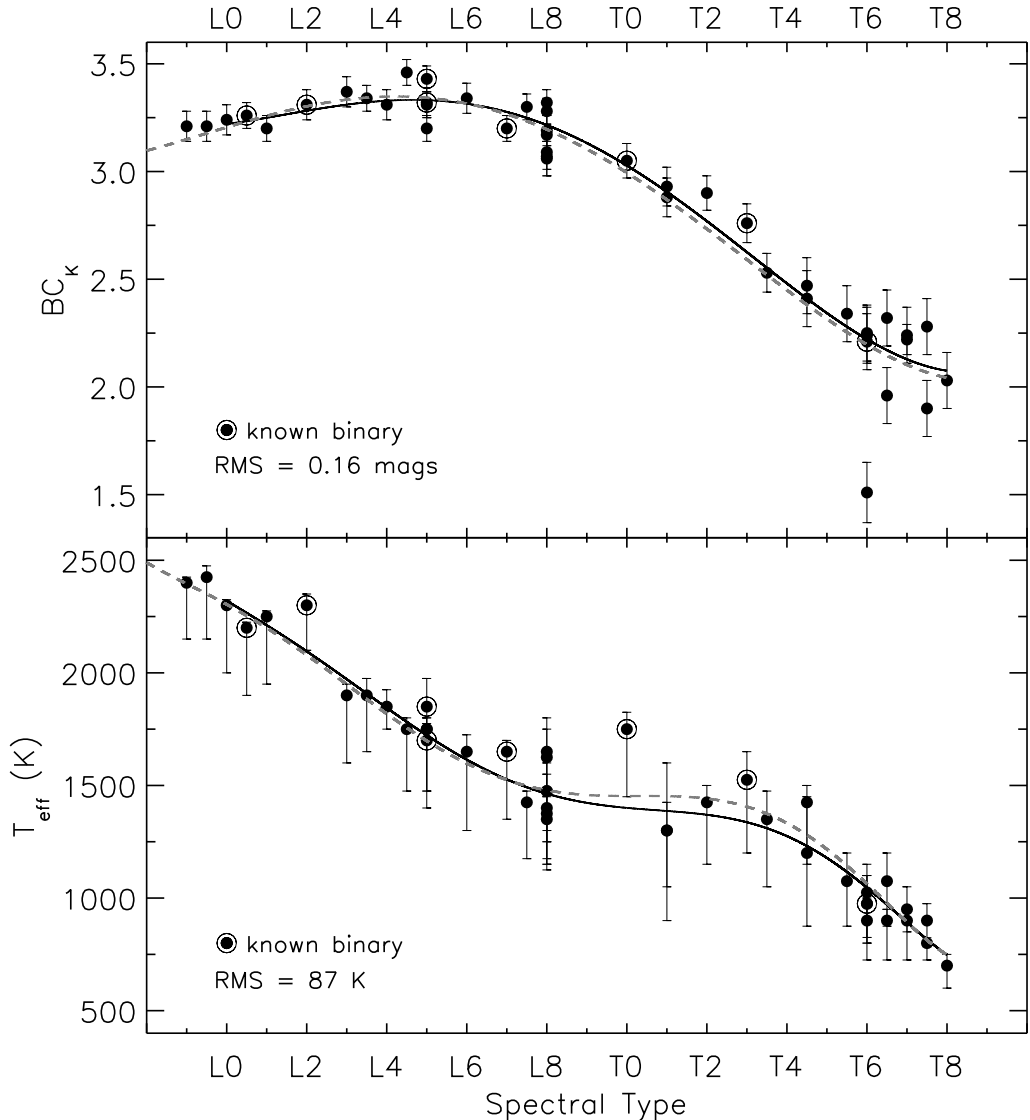}
\caption{Polynomial fits (solid lines) 
to M9--T8 objects with BC$_K$ measurements and
T$_{eff}$ estimates (from \citealt{2004AJ....128.1733G} and references
there-in), excluding known binaries. The relations are fourth- and
sixth-degree polynomial fits for BC$_K$ and T$_{eff}$, respectively,
with the coefficients listed in Table 3.  The plotted values used in the 
T$_{eff}$ relation are for an age of 3 Gyr, and the error bars represent
an age range of 0.1 -- 10 Gyr (younger ages are cooler and older ages
are hotter).  The BC$_K$-SpT relation is
weighted by the errors while the T$_{eff}$-SpT relation is unweighted.
The large outlier seen in the BC$_K$ plot and included in the fit 
is the T6 dwarf 2MASSI J0937347+293142, which is thought to be unusually 
blue due to high surface gravity/low-metallicity.  Shown for comparison
are the polynomial fits (dashed lines) from \cite{2004AJ....128.1733G}.
\label{fig8}}
\end{figure}

To estimate the physical parameters of 2MASS 1404AB (see Table 2), we 
first rederived MKO 
$K$-band bolometric corrections (BC$_K$) versus spectral type and T$_{eff}$
versus spectral type relations, using M9--T8 data from 
\cite{2004AJ....128.1733G}
and references there-in, excluding known binaries.  
In both relations, we classify L dwarfs on the
\cite{1999ApJ...519..802K} optical classification scheme, and T dwarfs on the \cite{2006ApJ...637.1067B} NIR classification scheme.  
Our choice of optical L dwarf spectral classification 
for these fits is further strengthened because
the far-optical spectrum of mid-to late-type L dwarfs is less influenced
by cloud opacities than the \cite{2002ApJ...564..466G} 1.5 $\mu$m H$_2$O 
index \citep{2004AJ....127.3553K}.  
The coefficients to the fourth-degree weighted BC$_K$ polynomial fit and 
to the sixth-degree unweighted T$_{eff}$ polynomial fit are given 
in Table 3 and shown
graphically in Fig.~\ref{fig8}.  The rms in this fit 
(87 K) compared to
that of \cite{2004AJ....128.1733G} (124 K) is less because of the 
elimination of
objects found to be binary between the time of that publication and this
paper.  Using the T$_{eff}$-SpT relation, we
estimate effective temperatures of $\sim$1390~$\pm$~90 K and 
$\sim$1180~$\pm$~90 K for the primary and secondary, respectively,
assuming an age of 3 Gyr.  We also find BC$_K$ values of
2.91~$\pm$~0.16 and 2.34~$\pm$~0.16 for the primary and secondary,
respectively.   In both cases, the error is the 
RMS in the fit and does not include the error in spectral typing.

To determine M$_{bol}$ for each component, we used the BC$_K$
corrections, a distance estimate
of d$\approx$23 pc, and the MKO component photometry (see Table 5).
This yielded M$_{bol}$~=~15.96~$\pm$~0.19 mag and 16.59~$\pm$~0.25 mag for
the primary and secondary, respectively.  The ratio of bolometric
luminosities, L$_{bol,A}$/L$_{bol,B}$, 
is 1.79~$\pm$~0.52, with the primary having the higher
luminosity.  We use the mass-luminosity relation
L$_{bol}$~$\propto$~M$^{2.64}$ \citep{2001RvMP...73..719B}, which 
assumes solar 
metallicities, and the ratio of bolometric luminosities to find the
mass-ratio: q~$\equiv$~M$_B$/M$_A$~=~0.80~$\pm$~0.09.  This mass
ratio is typical for known 
brown dwarf binaries, which tend to peak at q$\sim$1
(73\% of observed binaries have q~$\ge$~0.8, although, 
observational bias would favor high-q ratios; \citealt{2006ApJS..166..585B}). 
Using the age-luminosity
relation from \cite{2001RvMP...73..719B}, we find a total mass for the 
system of
50, 70, and 80 M$_{Jup}$, corresponding to ages of 0.5, 1.0, and 5.0 Gyr, 
typical for local disk dwarfs \citep{2000nlds.conf.....R}.  
This corresponds to individual
masses of $\sim$28 and 22 M$_{Jup}$, $\sim$39 and 31 M$_{Jup}$, 
and $\sim$44 and 36 M$_{Jup}$ for these respective ages.

Finally, this allows determination of the orbital parameters of this system.  
At an estimated distance of $\sim$23 pc and with 
an apparent separation of 0$\farcs$134, the 
projected physical separation is $\rho$~$\approx$~3.1 AU.  Statistically, the
true semi-major axis can then be estimated as 
a~=~1.26~$\times~\rho$~$\approx$~3.9 AU \citep{1992ApJ...396..178F}.  
This implies an
orbital period of roughly 28$-$35 yr, assuming a total system mass of 50$-$80 
M$_{Jup}$.  

\section{Conclusions}

We have presented high resolution imaging of two T2.5 dwarfs -- 
2MASS 1106 and 2MASS 1404 -- 
and resolved the latter into a 0$\farcs$13 binary.  2MASS 1404AB is an
intriguing binary as its presumably cooler secondary is brighter in
$J$-band than the primary by 2MASS $\Delta J$~=~0.45 mags but fainter in
both 2MASS $H$- (by 0.49 mags) and 2MASS $K_s$-bands (by 1.13 mags).  
This secondary brightening in $J$-band is much more pronounced 
than that seen in the other two known flux reversal binaries with resolved
$J$-band photometry -- SDSS 1021 (2MASS $\Delta J$~=~$-$0.05, MKO
$\Delta J$~=~0.04) and SDSS 1534 (2MASS $\Delta J$~=~0.03, MKO
$\Delta J$~=~0.17).  \cite{2007ApJ...666.1205Z} have found no companions 
in their radial velocity monitoring of the T4.5 dwarf 2MASS 0559, which 
lies $\sim$0.9 mag above the 2MASS M$_J$ versus spectral type relation 
we derive.  This result in conjunction with ours suggest that a
brightening of at least $\sim$0.5 mag and likely as high as $\sim$1 mag 
in the $J$-band bump is real and intrinsic to the atmospheres of
these objects as they cool across early-to mid-type T classes.  

\section{Acknowledgments}

We thank our anonymous referee for comments which improved this manuscript.  We would like to thank our telescope operators on Keck: Jim Lyke, Cynthia 
Wilburn, Chuck Sorenson, and the rest of the Keck LGS AO team and our
telescope operator on Magellan: Mauricio Martinez.  
We thank Michael Liu for the LGS AO data on these two sources, 
and Michael Cushing and Mark Pitts for helpful discussions.  
D.L.L.\ thanks Michael Liu and John Rayner for advising her for this
project, and David Sanders for financial support.  
This research has made use of the SIMBAD database,
operated at CDS, Strasbourg, France.  
This research has benefitted from the M, L, and T dwarf 
compendium housed at \url{DwarfArchives.org} and
has used data products from the SpeX Prism Spectral Libraries 
(\url{http://www.browndwarfs.org/spexprism}).
This publication also makes use of data products from the Two Micron All Sky 
Survey (2MASS), which is a joint project of the University of
Massachusetts and the Infrared 
Processing and Analysis Center/California Institute of Technology, 
funded by the National Aeronautics and Space Administration and the 
National Science Foundation.  This research has also made use of 
the NASA/IPAC Infrared Science Archive (IRSA), which is operated by the 
Jet Propulsion Laboratory, California Institute of Technology, under 
contract with the National Aeronautics and Space Administration.  As all 
data were obtained from the summit 
of Mauna Kea, the authors wish to recognize and acknowledge the very 
significant cultural role and reverence that this mountaintop has always 
had within the indigenous Hawaiian community. We are most fortunate to 
have the opportunity to conduct observations on the summit.

\noindent

\textbf{Facilities Used:} Keck:II, Magellan:Clay.

\begin{deluxetable}{lccccccc}
\tablewidth{6.8in}
\tablenum{1}
\tablecaption{$J$-Band Flux Reversal Binaries\tablenotemark{a}}
\tablehead{
\colhead{Designation} & \colhead{$J^{2MASS}$} & 
\colhead{$\Delta J_{A-B}^{2MASS}$\tablenotemark{b}} & 
\colhead{$\Delta J_{A-B}^{MKO}$\tablenotemark{c}} & 
\colhead{SpT AB\tablenotemark{d}} & \colhead{SpT A\tablenotemark{d}} &
\colhead{SpT B\tablenotemark{d}} & \colhead{Ref\tablenotemark{e}}} 
\startdata
SDSS 1021AB & 16.25~$\pm$~0.09 & $-$0.05 & 0.04 & T3 & 
  T1 & T5 & 1 \\
SDSS 1534AB   & 16.75~$\pm$~0.13 & 0.03 & 0.17 & T3.5 & 
  T1.5 & T5.5 & 2 \\
2MASS 1404AB  & 15.56~$\pm$~0.09 & 0.45 & 0.53 & T2.5 & 
  T1 & T5 & 3 \\
\enddata
\tablenotetext{a}{There is one other (possible) $J$-band flux reversal binary
not tabulated here: 2MASS 1728AB because it has been observed at $z$-band 
not at $J$-band (see $\S$1).}
\tablenotetext{b}{Relative 2MASS photometry calculated in this paper.}
\tablenotetext{c}{Relative MKO photometry from references.}
\tablenotetext{d}{NIR spectral types on \cite{2006ApJ...637.1067B} scheme.}
\tablenotetext{e}{References -- (1) \cite{2006ApJS..166..585B}, 
(2) \cite{2006ApJ...647.1393L}, (3) this paper.}
\end{deluxetable}

\begin{deluxetable}{lc}
\tablewidth{3.7in}
\tablenum{2}
\tablecaption{Properties of the 2MASS 1404AB System}
\tablehead{
\colhead{Property} & \colhead{Value}}
\startdata
MKO $J$ (mags) & 15.41~$\pm$~0.07 \\
MKO $H$ (mags) & 15.02~$\pm$~0.08 \\
MKO $K$ (mags) & 14.55~$\pm$~0.10 \\
MKO $\Delta J$ (mags) & 0.53~$\pm$~0.16 \\
MKO $\Delta H$ (mags) & $-$0.48~$\pm$~0.11 \\
MKO $\Delta K$ (mags) & $-$1.20~$\pm$~0.21 \\
2MASS $J$ (mags) & 15.58~$\pm$~0.07 \\
2MASS $H$ (mags) & 14.96~$\pm$~0.08 \\
2MASS $K_s$ (mags) & 14.54~$\pm$~0.10 \\
2MASS $\Delta J$ (mags) & 0.45~$\pm$~0.15 \\
2MASS $\Delta H$ (mags) & $-$0.49~$\pm$~0.13 \\
2MASS $\Delta K_s$ (mags) & $-$1.13~$\pm$~0.22 \\
Composite Optical SpT & T0 \\
Composite NIR SpT & T2.5\tablenotemark{a} \\ 
$\mu$ ($\arcsec$ yr$^{-1}$) & 0.35~$\pm$~0.03\tablenotemark{a} \\
$\theta$ (deg) & 275.3~$\pm$~0.2\tablenotemark{a} \\ 
$\rho$ (mas) & 133.6~$\pm$~0.6 \\
$\phi$ (deg) & 311.8~$\pm$~0.7 \\
log (L$_A$/L$_B$)\tablenotemark{b} & 0.25~$\pm$~0.13 \\
q~$\equiv$~M$_B$/M$_A$ & 0.80~$\pm$~0.09 \\
M$_{tot}$ (M$_{Jup}$) for 0.5, 1.0 \& 5.0 Gyr & 50, 70 \& 80 \\
Est$.$ distance (pc) & $\sim$23 \\
Est$.$ projected separation (AU) & $\sim$3.1 \\
Est$.$ actual separation (AU) & $\sim$3.9 \\
Orbital Period (yr) & 28~--~35 \\
\enddata
\tablenotetext{a}{Measurements are from \cite{2007AJ....134.1162L}.}
\tablenotetext{b}{$\frac{L_{bol}}{L_{bol\odot}}$~=~100$^{(M_{bol\odot}~-~~M_{bol})/5}$, where M$_{bol\odot}$~=~+4.76.} 
\end{deluxetable}

\begin{deluxetable}{lcccccccc}
\tabletypesize{\scriptsize}
\tablewidth{6in}
\tablenum{3}
\tablecaption{Coefficients of Polynomial Fits for L and T Dwarfs\tablenotemark{a}}
\tablehead{
\colhead{R} & \colhead{c$_0$} & \colhead{c$_1$} & \colhead{c$_2$} & 
\colhead{c$_3$} & \colhead{c$_4$} & \colhead{c$_5$} & \colhead{c$_6$} & 
\colhead{RMS}}
\startdata
M$_J$\tablenotemark{b}~~(mag) & 11.817 & 1.255e-1 & 3.690e-2 & 
  1.663e-2 & $-$3.915e-3 & 2.595e-4 & $-$5.462e-6 & 0.29 \\
M$_H$\tablenotemark{b}~~(mag) & 11.010 & 1.125e-1 & 3.032e-2 & 
  1.261e-2 & $-$2.970e-3 & 1.987e-4 & $-$4.218e-6 & 0.29 \\
M$_{K_s}$\tablenotemark{b}~~(mag) & 10.521 & 7.369e-2 & 2.565e-2 & 
  1.299e-2 & $-$2.864e-3 & 1.911e-4 & $-$4.104e-6 & 0.33 \\
BC$_K$\tablenotemark{c,d}~~~~(mag) & 3.221 & 2.371e$-$2 & 6.428e$-$3 & 
  $-$1.631e$-$3 & 5.579e$-$5 & \nodata & \nodata & 0.16 \\
T$_{eff}$\tablenotemark{d}~~(K) & 2319.920 & $-$108.094 & 1.950 & $-$3.101 
  & 6.414e$-$1 & $-$4.255e$-$2 & 9.084e$-$4 & 87 \\
\enddata
\tablenotetext{a}{Polynomial fits to optical L
dwarfs (classified on \citealt{1999ApJ...519..802K} scheme) and NIR T dwarfs
(classified on \citealt{2006ApJ...637.1067B} scheme) with parallax
measurements and not known to be binary (see $\S$3.1 \& $\S$3.4 for a 
full description).  Each function is
defined as R~=~$\displaystyle\sum_{i=0}^{n}~c_i~\times~(SpT)^i$
and is valid for spectral types L0$-$T8, where 0~=~L0, 10~=~T0, etc.  
These fits are shown graphically in Figs.~\ref{fig4} and ~\ref{fig8}.}
\tablenotetext{b}{Photometry is on the 2MASS system.}
\tablenotetext{c}{Photometry is on the MKO system.} 
\tablenotetext{d}{Data are from \cite{2004AJ....128.1733G} and references 
therein.}
\end{deluxetable}

\begin{deluxetable}{lcccccccc}
\tabletypesize{\scriptsize}
\tablewidth{6.8in}
\tablenum{4}
\tablecaption{$K$-band Differences Between 2MASS and MKO by Spectral Type:
L6.0--T2.0}
\tablehead{
\colhead{Quantity} & 
\colhead{L6} & \colhead{L7} & \colhead{L8} & \colhead{L9} & 
\colhead{T0} & \colhead{T1} & \colhead{T2} & \colhead{AVG}}
\startdata
2MASS $K_s$ $-$ MKO $K$ & 0.03~$\pm$~0.01 & 0.02 & 0.02~$\pm$~0.01 &
   0.01~$\pm$~0.01 & 0.01 & $-$0.01~$\pm$~0.02 & $-$0.03~$\pm$~0.02 & 
   $-$0.01~$\pm$~0.04 \\
Number of Objects\tablenotemark{a} & 4 & 1 & 2 & 2 & 1 & 7 & 8 & 25 \\
\enddata
\tablenotetext{a}{The number of objects available per spectral type
class, where a class also includes a half-subtype - i.e., L6.0 and L6.5.
Note that the T2 spectral class only includes T2.0 objects and no T2.5 objects.}
\end{deluxetable}

\begin{deluxetable}{lcc}
\tablewidth{4.2in}
\tablenum{5}
\tablecaption{Properties of the Components of the 
2MASS 1404AB System}
\tablehead{
\colhead{Property} & \colhead{2MASS 1404A} & \colhead{2MASS 1404B}}
\startdata
MKO $J$ (mags) & 16.46~$\pm$~0.12 & 15.93~$\pm$~0.09 \\
MKO $H$ (mags) & 15.56~$\pm$~0.09 & 16.04~$\pm$~0.10 \\
MKO $K$ (mags) & 14.86~$\pm$~0.11 & 16.06~$\pm$~0.19 \\
2MASS $J$ (mags) & 16.65~$\pm$~0.12 & 16.20~$\pm$~0.09 \\
2MASS $H$ (mags) & 15.51~$\pm$~0.09 & 16.00~$\pm$~0.10 \\
2MASS $K_s$ (mags) & 14.83~$\pm$~0.11 & 15.96~$\pm$~0.19 \\
MKO $J-H$ (mags) & 0.90~$\pm$~0.15 & $-$0.11~$\pm$~0.13 \\
MKO $H-K$ (mags) & 0.70~$\pm$~0.14 & $-$0.02~$\pm$~0.21 \\
MKO $J-K$ (mags) & 1.60~$\pm$~0.16 & $-$0.13~$\pm$~0.21 \\
2MASS $J-H$ (mags) & 1.14~$\pm$~0.15 & 0.20~$\pm$~0.13 \\
2MASS $H-K_s$ (mags) & 0.68~$\pm$~0.14 & 0.04~$\pm$~0.21 \\
2MASS $J-K_s$ (mags) & 1.82~$\pm$~0.16 & 0.24~$\pm$~0.21 \\
Est$.$ NIR SpT\tablenotemark{a} & T1~$\pm$~1 & T5~$\pm$~1 \\
Est$.$ T$_{eff}$ (K)\tablenotemark{b}  & 1390~$\pm$~90 & 1180~$\pm$~90 \\
M$_{bol}$ (mags) & 15.96~$\pm$~0.19 & 16.59~$\pm$~0.25 \\
\enddata
\tablenotetext{a}{Component NIR spectral types are derived in $\S$3.2.}
\tablenotetext{b}{Based on the T$_{eff}$ vs.\ SpT relation defined in
Table 3, using the component spectral types listed and assuming an age of 3 Gyr.}
\end{deluxetable}

\clearpage

 \end{document}